\documentclass[12pt]{revtex4}
\usepackage{ifpdf}
\usepackage[dvipdfm]{graphicx}
\usepackage{epsfig,amsbsy,bm,amssymb}
\usepackage{hyperref}
\usepackage{pslatex}

\newcommand{\isum}%
{\mathop{\hbox{$\displaystyle\sum\kern-13.2pt\int\kern1.5pt$}}}
\renewcommand{\r}{{\bm r}}

\newcommand{\p}{{\bm p}}

\newcommand{\bt}{\begin{tabular}}
\newcommand{\et}{\end{tabular}}

\newcommand{\eref}[1] {(\ref{#1})}
\newcommand{\Eref}[1] {Eq.~(\ref{#1})}

\newcommand{\Fref}[1] {Figure \ref{#1}}

\newcommand{\br}{\begin{eqnarray*}}
\newcommand{\er}{\end{eqnarray*}}
\newcommand{\ba}{\begin{eqnarray}}
\newcommand{\ea}{\end{eqnarray}}
\newcommand{\be}{\begin{equation}}
\newcommand{\ee}{\end{equation}}

\newcommand{\bp}{\begin{minipage}}
\newcommand{\ep}{\end{minipage}}

\begin{document}
\bibliographystyle{apsrev}

\title {On the origin of the cusp in the transverse momentum distribution for the process of strong field ionization.}

\author{I. A. Ivanov$^{1,2}$}
\email{Igor.Ivanov@anu.edu.au}

\affiliation{$^{1}$Center for Relativistic Laser Science, Institute for
Basic Science, Gwangju 500-712, Republic of Korea}

\affiliation{$^2$Research School of Physics and Engineering,
The Australian National University,
Canberra ACT 0200, Australia}

\date{\today}

\begin{abstract}
We study the origin of the cusp-structure in the transverse or lateral electron 
momentum distribution (TEMD) 
for the process of tunelling ionization driven by a 
linearly polarized laser pulse. We show that appearance of the cusp in the TEMD can
be explained as follows. Projection on the set of the Coulomb scattering states leads to appearance of 
"elementary" cusps which have simple structure as functions of the lateral momentum. 
This structure is independent of the detailed dynamics of the ionization process and can be 
described analytically. These "elementary" cusps can be used to describe the cusp-structure
in TEMD. 
\end{abstract}

\pacs{32.80.Rm 32.80.Fb 42.50.Hz}
\maketitle

\section{Introduction}

The seminal paper by Keldysh \cite{Keldysh64} laid out the distinction between 
tunelling and multi-photon regimes in the photo-ionization process. Particularly 
fruitful the Keldysh's paradigm proved for the study of the tunelling ionization, a photo-ionization process
characterized by the small values of the so-called Keldysh parameter $\gamma=\omega\sqrt{2|\varepsilon_0|}/E$
(here $\omega$, $E$ and $|\varepsilon_0|$ are the frequency, field strength and ionization potential
of the target system expressed in atomic units). 
Subsequent developments
\cite{Faisal73,Reiss80,ppt,adk1} elaborated on various aspects of this approach in the tunelling regime, making
it an extremely useful and versatile tool for understanding tunelling photo-ionization. Comprehensive reviews
of these developments (to which we will be referring below as tunelling theories) 
can be found in \cite{tunr,tunr2}. 

A remarkable feature of the tunelling regime is that 
one may still use to some extent the classical notions, such as that of the electron trajectory.
This fact has been extensively used in the modeling of the tunelling photo-ionization.
In this approach tunelling photo-ionization is regarded 
as a process in which electron first emerges into the continuum as a result of the
under-the-barrier tunelling.  This part of the problem is described quantum mechanically, producing 
probabilistic distributions of the electron's characteristics (typically velocities), which 
can be used as initial conditions for the subsequent classical modeling describing electron motion
after the ionization event. Such a procedure has been used with success to produce ionization spectra 
in good agreement with experiment \cite{tipis} for complicated systems where truly {\sl ab initio} treatment
becomes hopelessly complicated. 

These distributions which weight different initial conditions
in the electron's phase space are not themselves observed in the experiment. The electron momentum 
distribution measured at the detector can, however, provide an information about the distributions
at the moment of the ionization event, which offers an exciting possibility to look at this event 
experimentally \cite{Boge,cusp3}.  Of course, from
the strict quantum-mechanical point of view the  
notion of the electron escaping the atom at
a particular moment of time should be regarded with some caution \cite{itm4}), nevertheless, this picture
of electron escaping into the continuum proved extremely fruitful.

Tunelling theories predict simple Gaussian-like structures for these initial distributions. 
If the after-ionization-event motion is treated as guided by the
laser field only (ionic core potential is neglected), the momentum distributions at the detector retains
this Gaussian character, with a possible shift of the distribution in the momentum space due to the 
overall momentum electron acquires from the laser field after the ionization event \cite{tunr}. 
Of particular
interest, therefore, is the so-called transverse or lateral electron momentum distribution (TEMD), 
which describes the distribution of the electron momenta measured at the detector in the direction perpendicular to the 
polarization plane of the driving pulse. In the simple picture when electron motion
is guided by the laser field only, the TEMD is unaffected by the motion subsequent to the ionization event. 

This prediction is not always true. While the TEMD measured at the detector is a Gaussian 
for the driving pulse with close to circular polarization \cite{coul7}, it looks rather different
for the case of the linear polarization. It has been found \cite{cusp2} 
that for the case of the linearly polarized 
laser pulse the transverse  electron momentum distribution 
exhibits a sharp cusp-like peak at zero transverse momentum. 
We studied this transition from the
cusp-like to the Gaussian-like structure in TEMD
numerically using the {\it ab initio} solution of the time-dependent Schr\"odinger equation
(TDSE) in \cite{cuspm}. 

Study of the TEMD can provide other useful information. 
It has been demonstrated, both experimentally and theoretically \cite{cusp_all}, that 
the transverse electron momentum distributions in the
tunneling and over the barrier ionization regimes (OBI) evolve in markedly
different ways when the ellipticity parameter describing
polarization state of the driving laser pulse increases. This fact can be used to make a
a clear distinction between the tunneling and OBI regimes in
the experiment. 

In the present work we study the origin of the cusp-structure in TEMD for the case of the 
linearly polarized driving laser pulse in detail. In \cite{cusp2} this structure at
zero transverse momentum has been attributed to low-energy singularity of the Coulomb wave-function \cite{cusp2}.
We show that though this interpretation is basically correct there is more to the story.
Projection on the set of the Coulomb scattering states produces the 
"elementary" cusps which have simple structure as functions of the lateral momentum. 
This structure can be 
described analytically. These "elementary" cusps can be used to describe the cusp-structure
in TEMD.

\section{Theory and results}

We will be guided below to a considerable extent by the numerical results provided by the solution of the 
TDSE for a hydrogen atom. We will briefly describe the procedure, therefore.
We solve TDSE for a hydrogen atom in presence of a laser pulse: 
\begin{equation}
i {\partial \Psi(\r) \over \partial t}=
\left(\hat H_{\rm atom} + \hat H_{\rm int}(t)\right)
\Psi(\r) \ .
\label{tdse}
\end{equation}

Operator  $\hat H_{\rm int}(t)$ in \Eref{tdse} 
describes interaction of the atom with the
EM field. We use 
velocity form for this operator:

\be
\hat H_{\rm int}(t) = {\bm A}(t)\cdot \hat{\bm p}\ ,
\label{gauge}
\ee

with 

\be
{\bm A(t)}=-\int_{0}^{t}{\bm E(\tau)}\ d\tau\\.
\label{vp}
\ee

The laser pulse is linearly polarized along the $z$-direction, which we use as a quantization axis:

\be
E_z= E_0 f(t) \cos{\omega t} 
\label{ef}
\ee

For the base frequency of the pulse we use $\omega=0.057$ a.u.
(corresponding to the wavelength of 790 nm). 
The function  $f(t)= \sin^2(\pi t/ T_1)$ in \Eref{ef} (here $T_1=N T$ is a total pulse duration, 
$N$ is an integer, $T=2\pi/\omega$ is an optical cycle of the field). 
We report below results for various pulse durations $T_1$ and field strengths $E_0$.
TDSE is solved for a time interval $(0,T_1)$.  
Initial state of the system is a ground state of the hydrogen atom.

To solve the TDSE we employed the procedure described in the works 
\cite{dstrong,circ6}.
Solution of the TDSE is represented as a  series in spherical harmonics:
\be
\Psi({\bm r},t)=
\sum\limits_{l=0}^{L_{\rm max}}
f_{l}(r,t) Y_{l0}(\theta).
\label{basis}
\ee

The radial part of the TDSE is discretized on the grid with the step-size
$\delta r=0.1$ a.u. in a box of the size $R_{\rm max}=600$ a.u. We consider below
relatively short total pulse durations  and moderately strong field intensities (not exceeding
6 optical cycles and $3.5\times 10^{14}$ W/cm$^2$ respectively).  We used $L_{\rm max}=50$ in the
calculations reported below. 
The necessary checks ensuring that for such field parameters calculation is well converged with respect to 
$L_{\rm max}$ and $R_{\rm max}$ have been performed.

Substitution of the expansion \eref{basis} into the TDSE gives a 
system of coupled equations for the radial functions 
$f_{l}(r,t)$.
To solve this system 
we use the matrix
iteration method \cite{velocity1}.
Ionization amplitudes $a(\p)$ 
are obtained by projecting solution of the 
TDSE at the end of the laser pulse
on the set of the ingoing scattering states 
$\psi^{(-)}_{\p}(\r)$ of the hydrogen atom:

\be
\psi^{(-)}_{\p}(\r)=\sum_{l\mu} i^l e^{-i\eta_l(p)} Y^*_{l\mu}(\p)Y_{l\mu}(\r) R_{lp}(r)\ .
\label{in}
\ee

For the linearly polarized laser pulse and the coordinate system we employ
only the terms with $\mu=0$, of course, actually
contribute to the projection.
Differential photo-ionization cross-section is computed as
$P(\p)=|a(\p)|^2$. We are interested in the transverse or lateral electron
momentum distribution, 
describing probability to detect a photo-electron with a given value
of the momentum component $p_{\perp}$ perpendicular to the 
polarization plane. Because of the symmetry of the problem due to the linear polarization of the 
driving pulse any plane containing polarization vector can be chosen as a polarization plane.
Choosing $(y,z)$-plane as a polarization plane, we obtain for TEMD as function of the lateral
momentum $p_{\perp}=p_x$:

\be
W(p_{\perp})=\int P(p_x,p_y,p_z)\ dp_y\ dp_z
\label{wp}
\ee

\begin{figure}[h]
\begin{tabular}{cc}
\resizebox{80mm}{!}{\epsffile{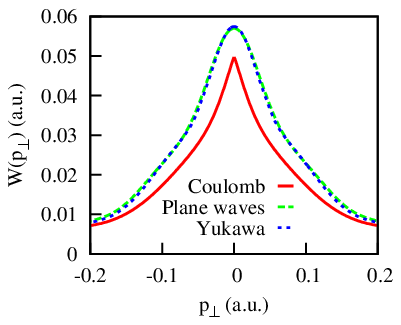}}  &
\resizebox{80mm}{!}{\epsffile{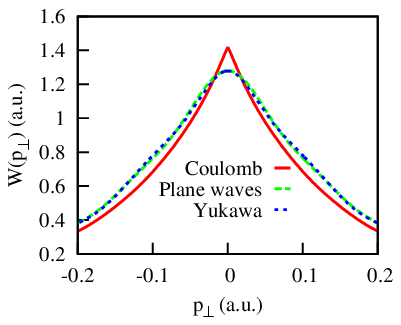}} \\ 
\end{tabular}
\caption{(Color online) Left panel: TEMD  
for the laser pulse \eref{ef} with pulse intensity of $10^{14}$ W/cm$^2$ and total duration
of 6 optical cycles. Right panel: the same for the field intensity of $3.5\times 10^{14}$ W/cm$^2$
and pulse duration of 4 optical cycles. Solid (red) line- projection on the ingoing Coulomb scattering states \eref{in}.
Dash (green)- projection on the basis of plane waves. Short dash (blue)- projection on the
set of the ingoing scattering states of the Yukawa potential.}
\label{fig1}
\end{figure}

TEMD obtained using this procedure are shown in
\Fref{fig1} for two sets of the driving pulse parameters.

Distributions obtained following the prescription described above and using the 
Coulomb ingoing scattering states \eref{in} for the projection operation (solid lines
in \Fref{fig1}) show the cusp-like behavior at $p_{\perp}=0$. While the TEMD
remains continuous at this point, its derivative suffers discontinuity. 
It was suggested in 
\cite{cusp2} that cusp originates
from the singularity of the Coulomb continuum
wave-function at zero energy. We subsequently found some numerical evidence \cite{cuspm} 
supporting this statement. 
What interests us in the present work is 
elucidating the nature of the cusp and the precise type of the discontinuity which the 
lateral distribution has at $p_{\perp}=0$. As we shall see, some analytical results describing
the discontinuity can be obtained. 

We begin by presenting results of a few numerical experiments. We make sure first that
the  cusp is indeed due to the projection on the set of the Coulomb continuum
wave-functions. In \Fref{fig1} we present results obtained if the same solutions of the TDSE
at the end of the laser pulse are projected on the set of the plane-waves and the ingoing states
of the Yukawa potential $V(r)=-e^{-0.1r}/r$ instead of the Coulomb scattering states. 
The spectra obtained by using plane-waves basis and 
scattering stated of the Yukawa potential, though agreeing quantitatively rather well with the 
spectrum obtained by projecting solution of the TDSE on the Coulomb scattering states, show no
cusp-like behavior at $p_{\perp}=0$. The cusp arises, therefore, as a result of the projection 
operation using the Coulomb scattering states as was surmised in \cite{cusp2}.
The question which
interests us is the detailed mechanism responsible for the appearance of the cusp. 

We note first that cusp cannot be introduced by the
integration procedure, when overlaps between the TDSE solution and the Coulomb scattering states are computed.
The amplitude functions $f_{l}(r,T_1)$ in  the expansion of the TDSE solution \eref{basis} 
are square-integrable functions with typical spatial extent 
corresponding to the distance the outgoing electron wave-packet can have traveled by the end of the laser pulse.
Integration of such functions cannot introduce any low-energy singular behavior. Indeed, we can consider
that to a good approximation the amplitude functions $f_{l}(r,T_1)$ have finite support, being non-zero only in
the finite region of space (this is what they are in the numerical calculation anyway). Integration of such functions
cannot introduce any new singularities which are not already present in the integrand. We must, therefore, look
carefully at the singularities present in the Coulomb scattering state \eref{in}.

There are two factors in \Eref{in} we have to examine: the Coulomb scattering phase-shifts and
the radial Coulomb wave-functions.
Explicit expression for the Coulomb phase-shifts reads \cite{LL3}:

\be
\eta_l(p)=\arg{{\rm\Gamma}\left(l+1-{i\over p}\right)} \ ,
\label{pf}
\ee

and it exhibits a highly singular behavior at $p=0$. 
On the other hand, the radial functions $R_{lp}(r)$ 
in \Eref{in} can be written (we use the $\delta(\p-\p')$ normalization) as \cite{LL3}:

\be
R_{lp}(r)= \beta(p)\gamma(p) g_{lp}(r) \ ,
\label{rr}
\ee

where

\be
\gamma(p)={1\over 1-e^{-{2\pi\over p}}} \ ,
\label{gam}
\ee

and

\be
\beta(p)={1\over\sqrt{p}} \ .
\label{bet}
\ee

The function $g_{lp}(r)$ can be found as the solution of the radial Schr\"odinger equation 
satisfying a boundary condition $\displaystyle g_{lp}(r)\to C_l e^{l+1}$ when $r\to 0$ and 
$C_l$ is a constant factor independent of energy. By the well-known Poincare theorem 
$g_{lp}(r)$ is, therefore, an entire function of energy, i.e. an entire function of $p^2$. 
The radial wave-function in \eref{in} is singular at $p=0$ only due to the presence of the factors 
$\gamma(p)$ and $\beta(p)$ given by \Eref{gam} and \Eref{bet}. 

We have identified, thus, three potential culprits which may introduce singular low-energy behavior and which 
may be responsible for the formation of a cusp. Let us study them one by one.

\begin{figure}[h]
\resizebox{120mm}{!}{\epsffile{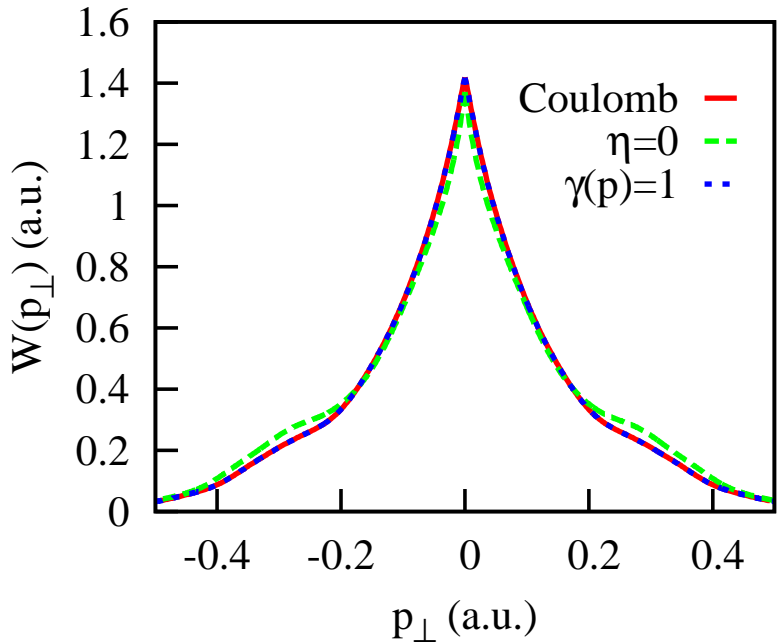}} 
\caption{(Color online) Left panel: TEMD 
for the laser pulse \eref{ef} with pulse intensity of $3.5\times 10^{14}$ W/cm$^2$
and pulse duration of 4 optical cycles. Solid (red) line- projection on the correct ingoing Coulomb states \eref{in}.
Dash (green)- projection on the set of states \eref{in} with $\eta_l(p)=0$.
Short dash (blue)- projection on set of the states \eref{in} with $\gamma(p)=0$.}
\label{fig2}
\end{figure}

Consider first the effect of the Coulomb scattering phase-shift. In \Fref{fig2} we present 
results of a simple numerical experiment obtained if in the expression for the Coulomb ingoing 
scattering state \eref{in} we put $\eta_l(p)=0$ in the exponential function (and use 
the correct Coulomb radial wave-functions $R_{lp}(r)$).

One can see that removal of the scattering phase-shifts $\eta_l(p)$ form the \Eref{in} hardly produces any effect
on the lateral spectrum. We could, in fact, anticipate this. Indeed, from the expression for the Coulomb
phase-shifts \eref{pf} and elementary properties of the Gamma-function one can easily deduce the relation:
$\displaystyle \eta_{l+1}(p)=\eta_l(p)-{\pi\over 2} + O(p) $ and hence:

\be
\eta_{l+1}(p)=\eta_0(p)-{(l+1)\pi\over 2} + O(p)\ ,
\label{pf1}
\ee

valid when $p\to 0$. We see, thus, that at low energies the effect of the Coulomb phase-shifts $\eta_{l}(p)$
in \Eref{in} reduces to introducing an energy-independent phase-factors for different terms in the sum in \Eref{in}, 
and overall dependence of the photo-ionization 
amplitude on the $\eta_0(p)$, which cancels out when we compute the squared modulus of the amplitude. On the other hand, 
with increasing energy Coulomb phase-shifts decrease fast, which explains the fact that Coulomb phase-shifts have virtually
no effect on the TEMD. The role of the factor $\gamma(k)$ is equally insignificant as can be seen from \Fref{fig2}, 
where we show the spectrum obtained if we put $\gamma(p)=1$ in the \Eref{in}. Again, this fact could be anticipated, since
this factor, though singular at $p=0$, does not blow up at this point and tends to be one with increasing energy. 

We are left, therefore, with the only factor in the \Eref{in} which blows up at $p=0$, the 
factor $\beta(p)$ in \Eref{bet}. There are, of course, other factors in \Eref{in} which are singular
at $p=0$. The spherical harmonics $Y_{l\mu}(\p)$ are, strictly speaking, singular functions of the components of the
vector $\p$ at $\p=0$ because of the coordinate system singularity at this point. Similarly, the higher order terms
in \Eref{pf1} are proportional to $p$, and, in a strict mathematical sense the function $p$ is singular at $\p=0$ 
as a function of the components of the vector $\p$. These singularities are, however, only mild ones, in particular
they do not lead to the unbounded growth of the function. The only singularity which does lead to such a growth is 
the one due to the factor $\beta(p)$. 

We are now in a position to elucidate the nature of the cusp in the 
lateral distribution. To this end, let us note that because of the symmetry of the problem the differential 
photo-ionization cross-section $P(\p)$ is, in fact, a function of two variables only:
$P(\p)=P(p,\cos\theta)$, where $\theta$ is the angle between electron momentum $\p$ and the $z$-axis.
Expanding this expression in powers of $\theta$ we may write for the differential cross-section:

\be
P(\p)=\sum\limits_{n=0}^{\infty} P_n(p)\cos^n{\theta} \ ,
\label{pn}
\ee 

where $P_n(p)$ are functions of of $p$ only. Coefficients of this expansion can
be computed numerically from the known solution of the TDSE by 
re-expanding products of the spherical harmonics $Y_{l\mu}(\p)$ occurring in
the expression for the squared modulus of the amplitude $|a(\p)|^2$ in 
series of spherical harmonics with the help of the well-known formulas, and re-expanding in turn 
the resulting  spherical
harmonics in powers of $cos{\theta}$. Important point here is that coefficients $P_n(p)$ depend only 
on $p$ and inherit from the amplitudes the singular behavior at $p=0$ due to the factor $\beta(p)$ in \Eref{bet}. 
The $p^{-{1\over 2}}$ singular 
behavior of the amplitudes at $\p=0$ clearly entails the $p^{-1}$ singular behavior of the coefficients 
$P_n(p)$ at $p=0$. Integrating \Eref{pn} over the $(p_y,p_z)$-plane (only terms with even $n$ 
give nonzero contributions, of course) we obtain for the TEMD:

\be
W(p_{\perp})=\sum\limits_{n=0,2,...}^{\infty} W_n(p_{\perp}) \ ,
\label{pn1}
\ee 

where

\be
W_n(p_{\perp}) = \int\limits_0^{\infty}\ dq\int\limits_0^{2\pi} \ d\phi
q P_n(p) \left({q\cos{\phi}\over p}\right)^n = {2\pi n!\over {\rm \Gamma}(n/2+1)^2 2^{n-1}}
 \int\limits_0^{\infty}\ dq P_n(p) {q^{n+1}\over \left(p_{\perp}^2+q^2\right)^{n\over 2} } \ ,
\label{pn2}
\ee 

where $p=\sqrt{p_{\perp}^2+q^2}$, and we used a cylindrical coordinate system $(q,\phi)$ in the 
$(p_y,p_z)$-plane.

Using \Eref{pn2} we can obtain asymptotic behavior of $W_n(p_{\perp})$ for $p_{\perp}\to 0$. As we mentioned
above $P_n(p)$ behave as $p^{-1}$ for small energies. Let us choose some 
small positive $Q$ and represent $P_n(p)$ in the interval $(0,Q)$ as $P_n(p)= C_n/p + P'_n(p)$ where $P'_n(p)$
is non-singular at $p=0$. Singular behavior of the integrals in \Eref{pn2} 
at  $p_{\perp} = 0$ can appear only as a result of the integration of the 
singular term $C_n/p$ in $P_n(p)$ over the interval $(0,Q)$. Indeed, the integrands in both integral over
the interval $(0,\infty)$ with $P'_n(p)$ and the integral over the interval $(Q,\infty)$ 
with $C_n/p$ contain smooth regular functions
as integrands which cannot lead to a small-$p_{\perp}$ singular behavior. 
We can write, therefore: 

\be
W_n(p_{\perp}) = C_n I_n(p_{\perp},Q) + W^{\rm reg}_n(p_{\perp})  \ ,
\label{pn3}
\ee

where 

\be
I_n(p_{\perp},Q)=\int\limits_0^{Q}\ dq {q^{n+1}\over \left(p_{\perp}^2+q^2\right)^{n+1\over 2}} \ ,
\label{pn33}
\ee 

$C_n$ is a constant, and $W^{\rm reg}_n(p_{\perp})$ are non-singular at $p_{\perp} = 0$, behaving near this point as
$W^{\rm reg}_n(p_{\perp})\approx u_n^0+ u_n^2p_{\perp}^2$ with some constant $u_n^0$ and $u_n^2$.

It is clear from \Eref{pn33} why this contribution becomes singular at $p_{\perp}=0$. While the integral has a finite
value if we put $p_{\perp}=0$, an attempt to calculate the derivative with respect to $p_{\perp}$ by differentiating under the
integral sign leads to a divergent expression. We have to be more careful in evaluating asymptotic of the integral.
Using well-known formulas for the integral representation and asymptotic properties of the hypergeometric function
$F(a,b;c;z)$ \cite{abr} we obtain:

\be
I_n(p_{\perp},Q) = { Q^{n+2}\over 2|p_{\perp}|^{n+1}} 
{{\rm \Gamma}\left({n\over 2}+1\right) {\rm \Gamma}\left({1\over 2}\right)\over {\rm \Gamma}\left({n\over 2}+{3\over 2}\right)}
F\left( {n+1\over 2}, {n\over 2}+1; {n\over 2}+{3\over 2}; -{Q^2\over p_{\perp}^2}\right) \approx 
A_n+ B_n |p_{\perp}| \qquad p_{\perp}\to 0 \ ,
\label{pn4}
\ee

where we do not give explicit expressions for the unimportant constant factors.

\begin{figure}[h]
\begin{tabular}{cc}
\resizebox{80mm}{!}{\epsffile{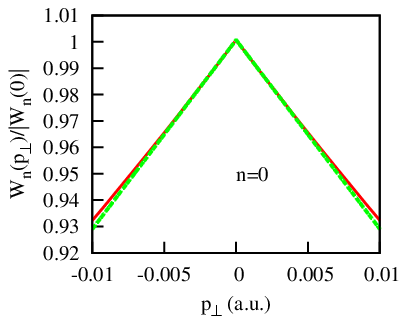}}  &
\resizebox{80mm}{!}{\epsffile{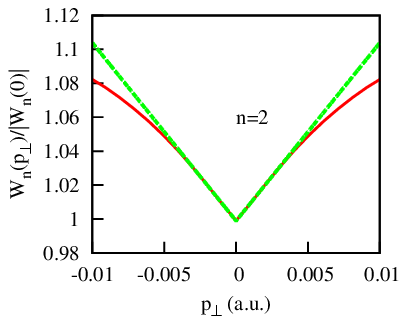}} \\ 
\resizebox{80mm}{!}{\epsffile{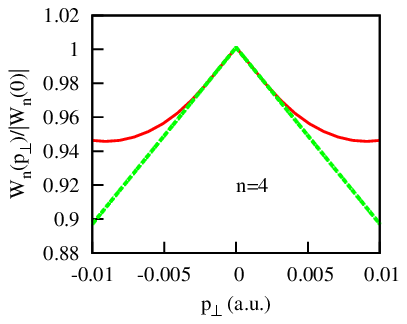}}  &
\resizebox{80mm}{!}{\epsffile{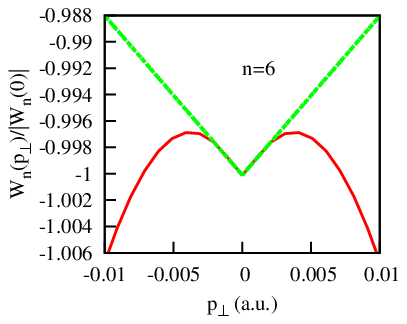}} \\ 
\end{tabular}
\caption{(Color online). Solid (red) line: scaled terms of the series \eref{pn1}  $W_n(p_{\perp})/|W_n(0)|$ as functions of lateral momentum 
$p_{\perp}$. Dash (green): linear fit: $W_n(p_{\perp})/|W_n(0)|= A+B|p_{\perp)}|$.
Field intensity $3.5\times 10^{14}$ W/cm$^2$, 
pulse duration four optical cycles.}
\label{fig3}
\end{figure}

From \Eref{pn3} and \Eref{pn4} we conclude that $W_n(p_{\perp})$ behave for small $p_{\perp}$ as linear functions
of $|p_{\perp}|$:

\be
W_n(p_{\perp},Q) = g_n+ h_n |p_{\perp}| + O\left(p_{\perp}^2\right) \qquad p_{\perp}\to 0 \ ,
\label{pn44}
\ee 

where $g_n$, $h_n$ are some constants. This implies the following
cusp structure of  $W_n(p_{\perp})$ at $p_{\perp}=0$. First derivative of $W_n(p_{\perp})$ with respect
to $p_{\perp}$ is discontinuous at $p_{\perp}=0$, the second derivative is, therefore, infinite.
That this formula
indeed reproduces asymptotic behavior correctly, can be seen from \Fref{fig3}. 
The contributions $W_n(p_{\perp})$ as functions of lateral momentum obtained from the TDSE calculation are shown in \Fref{fig3}. 
$W_n(p_{\perp})$ vary considerably with $n$ in magnitude and need not be positive, to facilitate the 
comparison we present the scaled contributions: $W_n(p_{\perp})/|W_n(0)|$. 
In \Fref{fig3} we present also the results of the linear fits: $W_n(p_{\perp})/|W_n(0)|= A+B|p_{\perp)}|$.

\begin{figure}[h]
\begin{tabular}{cc}
\resizebox{80mm}{!}{\epsffile{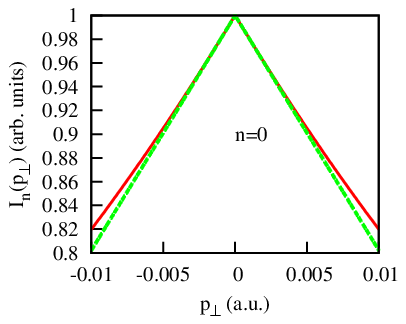}}  &
\resizebox{80mm}{!}{\epsffile{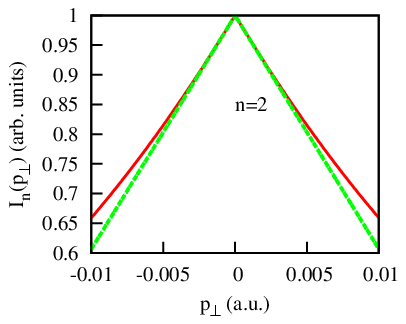}} \\ 
\resizebox{80mm}{!}{\epsffile{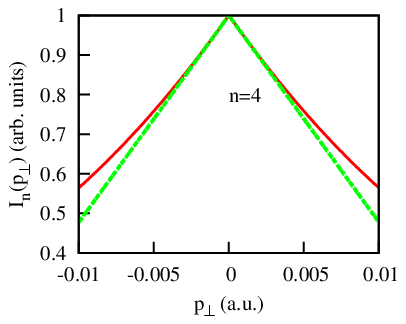}}  &
\resizebox{80mm}{!}{\epsffile{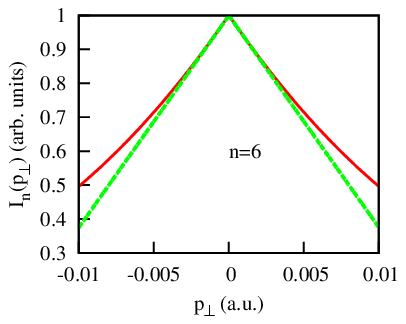}} \\ 
\end{tabular}
\caption{(Color online) 
Solid (red) line: Integrals $I_n(p_{\perp},Q)$ as functions of lateral momentum 
$p_{\perp}$ for $Q=0.05$.  Dash (green): linear fit: $I_n(p_{\perp},Q)= A+B|p_{\perp)}|$.}
\label{fig4}
\end{figure}

Two features are apparent from
\Fref{fig3}. First, for small values of $p_{\perp}$ the contributions $W_n(p_{\perp})$ are indeed linear 
functions of $|p_{\perp}|$ in agreement with the asymptotic estimate \eref{pn44} we made above. Second, 
the region where this asymptotic estimate represents $W_n(p_{\perp})$ accurately shrinks with $n$. 
There is, of course,  nothing unusual in such behavior. Asymptotic estimates give us asymptotic
behavior for small values of a parameter, but they do not necessarily tell us how small the parameter should be 
for the estimate to be accurate.  A glance at the behavior of the integrals in \Eref{pn33} as functions of
lateral momentum may help us to understand what is happening. We show in \Fref{fig4} integrals 
$I_n(p_{\perp},Q)$ as functions of $p_{\perp}$ for a fixed value of $Q=0.05$ a.u. \Fref{fig4} shows
qualitatively the behavior similar to what we observe in \Fref{fig3}, the region where the asymptotic expression
represents $I_n(p_{\perp},Q)$ accurately shrinks with $n$. 
Thus, the behavior of $W_n(p_{\perp})$ is a
consequence of the property of the integrals $I_n(p_{\perp},Q)$ that the region where linear in $|p_{\perp}|$
asymptotic takes over shrinks progressively with $n$. 

This feature of $W_n(p_{\perp})$ is quite important
for understanding  how the cusp in TEMD is produced. 
As \Eref{pn44} shows the terms of the series \eref{pn1} behave as linear functions of 
$|p_{\perp}|$ for small enough $p_{\perp}$. At first glance, that would suggest that the sum of the series \eref{pn1}, the
TEMD $W(p_{\perp})$, would exhibit the same behavior linear in $|p_{\perp}|$ near $p_{\perp}=0$. That would imply
the following cusp structure: TEMD would have discontinuous first and infinite second derivative at
$p_{\perp}=0$.  On the other hand, the cusps shown in \Fref{fig1} look more like functions of $p_{\perp}$ with 
infinite first derivative. In other words $W(p_{\perp})$ grows visibly faster than $|p_{\perp}|$ near $p_{\perp}=0$.
This apparent contradiction is resolved when one realizes that $W(p_{\perp})$ is a sum \eref{pn1}  of the terms 
$W_n(p_{\perp})$. Each of $W_n(p_{\perp})$ behaves as a linear function of $|p_{\perp}|$ in some vicinity of 
$p_{\perp}=0$, but, as we saw above, the interval of $p_{\perp}$ on which linear dependence is a good approximation shrinks
with $n$. If, as \Fref{fig3} suggests, outside the interval of applicability of the asymptotic law the 
$W_n(p_{\perp})$ grow slower with $p_{\perp}$ (we see this behavior for the integrals $I_n(p_{\perp},Q)$ 
in \Fref{fig4}), the sum of all $W_n(p_{\perp})$ in \eref{pn1} will exhibit precisely the behavior seen in \Fref{fig1}- the 
growth which is faster than linear for $p_{\perp} \to 0$. With $p_{\perp} \to 0$ the terms
with higher $n$ switch progressively from the relatively slow growth outside the asymptotic region to 
a faster growth linear in $|p_{\perp}|$, once $p_{\perp}$ is inside the region of the validity of the asymptotic law
\eref{pn44} for a particular $n$. 

To see this quantitatively we introduce a function:

\be
f(z)= \sum_{m=0}^{\infty} C_{2m}z^m\ ,
\label{ff}
\ee 

where $C_{2m}$ are the coefficients in \Eref{pn3} (the definition \eref{ff} takes into account that
only the even-order coefficients $C_n$ occur in \Eref{pn1}). The function $f(z)$ in \Eref{ff} is defined 
in terms of the set of the coefficients $C_n$ and encapsulates, therefore, information about the 
solution of the TDSE and, ultimately, information about the dynamics of the ionization process. 
For the sum of the series \eref{pn1} we can write then (we omit the contributions of the regular parts
$W^{\rm reg}_n(p_{\perp})$ since they do not lead to the singular behavior):

\be
I(p_{\perp})= \sum\limits_{n=0,2,...}^{\infty}C_n \int\limits_0^{Q}\ dq {q^{n+1}\over \left(p_{\perp}^2+q^2\right)^{n+1\over 2}}=
\int\limits_0^{Q} {q\over \sqrt{p_{\perp}^2+q^2}}f\left({q^2\over p_{\perp}^2+q^2} \right)  \ dq \ .
\label{ex}
\ee

Substituting $\displaystyle y= {p_{\perp}\over \sqrt{p_{\perp}^2+q^2}} $ we can rewrite
\Eref{ex} as:

\be
I(p_{\perp})= p_{\perp} \int\limits_{y(Q)}^1 {dy\over y^2} f\left(1-y^2\right),
\label{tt}
\ee

where $\displaystyle y(Q)= {p_{\perp}\over \sqrt{p_{\perp}^2+Q^2}} $.

Assuming that in this expression $f(z)=z^n$, which corresponds to all but one coefficients $C_n$ in \Eref{pn3}
having zero values, reproduces, of course,  the asymptotic law \eref{pn44} we
obtained above for the individual terms $W_n(p_{\perp})$ of the series \eref{pn1}. More realistic 
assumption about $f(z)$ can be based on the observation we made above that
the coefficients $C_n$ are ultimately related to the partial wave expansion of the solution of the TDSE equation.
It is clear that to establish
small-$p_{\perp}$ behavior of the sum of the series \eref{pn1} we actually need to know only the large-$n$ asymptotic
behavior of $C_n$.
Partial wave expansions in the TDSE calculations converge, as a rule, rather slowly \cite{velocity1}.  
A plausible assumption about the  large-$n$ asymptotic behavior of the sequence of $C_n$, reflecting this
slow convergence, would be 
a power-like asymptotic behavior $C_n \propto n^{-\lambda}$ with some positive $\lambda$. By the well-known theorems
of the complex analysis this implies presence of at least one 
singular point of $f(z)$ on the circle of convergence $|z|=1$. Let us assume, for
example, that this singularity is a simple branch point at $z=1$, so that $\displaystyle f(z)=\sqrt{1-z}$.
We obtain then from \Eref{tt}:

\be
I(p_{\perp})= p_{\perp} \int\limits_{y(Q)}^1 {dy\over y} \ ,
\label{tt}
\ee

which, as one can easily see, leads to the following formula for the 
asymptotic behavior of $I(p_{\perp})$ for $p_{\perp}\to 0$:

\be
I(p_{\perp}) = A + B|p_{\perp}|\ln{|p_{\perp}|} \qquad p_{\perp}\to 0 \ ,
\label{pn5}
\ee 
  
with some constants $A$ and $B$. We see, thus, that while terms of the series \eref{ex} all
have $|p_{\perp}|$-cusps at $p_{\perp}=0$, the sum of the series \eref{ex} can exhibit
more singular cusp-like behavior at this point.
For the terms of series \eref{ex} the first derivatives  with respect
to $p_{\perp}$ are discontinuous at $p_{\perp}=0$ and the second derivative is infinite at $p_{\perp}=0$.
The cusp-singularity of the sum of the series \eref{ex}, with the choice of the function $f(z)$ we used above as an example,
is more severe, the first derivative 
with respect to $p_{\perp}$  is infinite at $p_{\perp}=0$.

\begin{figure}[h]
\resizebox{120mm}{!}{\epsffile{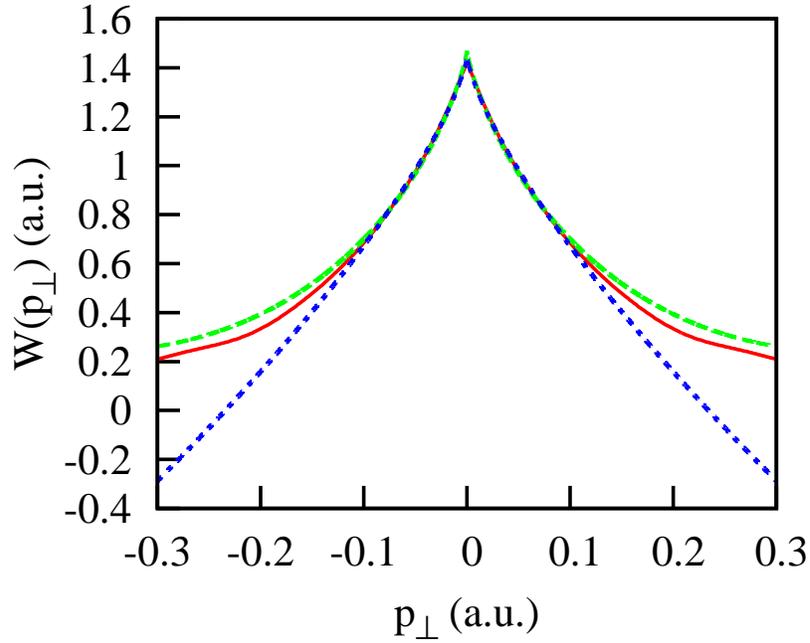}}  
\caption{(Color online) TEMD for 
the field intensity of $3.5\times 10^{14}$ W/cm$^2$
and pulse duration of 4 optical cycles. Solid (red) line: TDSE calculation.
Dash (green): fit based on the equation:  $W(p_{\perp})=A + B|p_{\perp}|\ln{|p_{\perp}|}$ 
(fitting parameters $A$ and $B$).
Short dash (blue): fit based on the equation:  $W(p_{\perp})=A + B|p_{\perp}|^{\alpha}$
(fitting parameters $A$, $B$ and $\alpha$).}
\label{fig6}
\end{figure}

Incidentally, the asymptotic law \eref{pn5} reproduces fairly well the TEMD
obtained in our TDSE calculations.
As one can see from \Fref{fig6}, the two-parameter formula $A + B|p_{\perp}|\ln{|p_{\perp}|}$ 
($A$ and $B$ considered as fitting parameters) gives
actually better results than the three-parameter fit based on the equation:
$W(p_{\perp})=A + B|p_{\perp}|^{\alpha}$ ($A$, $B$ and $\alpha$ as fitting parameters). 
We cannot, of course, claim that $A + B|p_{\perp}|\ln{|p_{\perp}|}$
is the actual behavior of the TEMD $W(p_{\perp})$ for small $p_{\perp}$. As we saw, to describe
the cusp in the TEMD we rely on two ingredients. We need first to describe small-$p_{\perp}$ behavior of the 
"partial" distributions $W_n(p_{\perp})$ in the \eref{pn1}. \Eref{pn3} and \Eref{pn4} provide an answer to this problem.
To find small-$p_{\perp}$ behavior of the sum of the series \eref{pn1} we also need to know the weights with which 
different  $I_n(p_{\perp})$ in \Eref{pn3} contribute to the sum- the coefficients $C_n$ in this equation. 
The logarithmic behavior in \Eref{pn5} obtains, in particular, assuming that $f(z)$ in 
\Eref{ff}, which encapsulates information about the coefficients $C_n$, 
has a simple branch point at $z=1$, an assumption leading to apparently satisfactory results,
which we, however,
did not prove. We may regard this formula, therefore, as a plausible but only a tentative expression.

\section{Conclusion}

We considered in detail the formation of the cusp in the TEMD for the process of 
strong field ionization. 
As we saw, one can push analytic approach to this problem quite far. Our starting point was
the series \eref{pn1}
resulting from the expansion of the differential probability in the powers of $\cos{\theta}$- the angle between
the polarization vector and the electron momentum. We were able to show that the terms $W_n(p_{\perp})$ of this
series behave as linear functions of  $|p_{\perp}|$ for small $p_{\perp}$. This behavior is a consequence of 
the properties of the Coulomb continuous spectrum wave-functions, and is present, therefore, for any system
regardless of what the actual Hamiltonian is, as long as the wave-function of the system after the end of the laser pulse 
is projected on the set of the 
Coulomb wave-functions. If this were the whole story the TEMD
would have the $|p_{\perp}|$-cusp at $p_{\perp}=0$. For such a cusp the first derivative at 
$p_{\perp}=0$ is discontinuous, and the second derivative is infinite.  

There is, however, the second step we have to perform to obtain the TEMD.
The functions $W_n(p_{\perp})$ in \eref{pn1} with the small-$p_{\perp}$ asymptotic which we
established in \Eref{pn44}, constitute  the building blocks from which TEMD can 
be build by summing up the expansion \eref{pn1}. 
It is at this stage, where dynamic information, i.e. the information about 
particular details of the ionization process, becomes important. 
The "partial" lateral distributions  $W_n(p_{\perp})$ considered as functions of $n$ 
enter the series \eref{pn1} with different weights. Mathematically, it is reflected in the \Eref{pn3} 
which represents $W_n(p_{\perp})$ as a product of the integral $I_n(p_{\perp})$
and a coefficient $C_n$ which is a function of $n$ only. Coefficients $C_n$
depend, of course, on the dynamics of the system, since they result ultimately from the 
projection of the TDSE wave-function at the end of the laser pulse.
The function we introduced 
in \Eref{ff} conveniently encapsulates this information. 
As we saw, the summation procedure 
can make the character of the cusp for the sum of the series \eref{pn1} different from the 
linear $|p_{\perp}|$-cusp which each of the terms of the series exhibits at $p_{\perp}=0$.
The reason for this is, roughly speaking,  the fact that for the terms of the series \eref{pn1}
with higher $n$ the region of lateral momenta for which linear in $|p_{\perp}|$ asymptotic law holds for 
$W_n(p_{\perp})$ shrinks, or in stricter mathematical language, the fact that the small-$p_{\perp}$ 
asymptotic \eref{pn44} for $W_n(p_{\perp})$ is non-uniform in $n$.

To summarize, we demonstrated that the cusp in the TEMD arises as a consequence of two factors:
The singularity of the Coulomb wave-function produces a simple cusp of the $A+B p_{\perp}$ type. 
The view expressed in \cite{cusp2} that cusp is due to the singularity of the Coulomb scattering state is,
therefore, basically correct, the Coulomb wave-function is responsible for the presence of the cusp.
This fact has nothing to do whatsoever with dynamics of the photo-ionization process. 
The character of  the cusp we observe in the TEMD, however, may differ
from the $A+B p_{\perp}$ -type created by the Coulomb wave-function. The origin of this difference
lies in the dynamics, it is ultimately due to the properties of the coefficients of the expansions \eref{pn},
\eref{pn1}, which depend on the wave-function at the end of the laser pulse.

\section{Acknowledgments} 

This work was supported by the Institute for Basic Science
under IBS-R012-D1.


\end{document}